\documentclass{pasj02}

\usepackage[whole]{bxcjkjatype}
\usepackage{natbib}
\usepackage{url}

\def\arcdeg{$^\circ$}

\Received{2025/12/19}
\Accepted{2026/02/03}
\Published{yyyy/mm/dd}

\begin{document} 

\title{Kinematic-Distance Biases in the Inner Milky Way from a Stellar-Dynamically Constrained Bar}

\author{Junichi \textsc{Baba}\altaffilmark{1,2}}
\email{babajn2000@gmail.com; junichi.baba@sci.kagoshima-u.ac.jp}

\altaffiltext{1}{Amanogawa Galaxy Astronomy Research Center, Graduate School of Science and Engineering, Kagoshima University, 1-21-35 Korimoto, Kagoshima 890-0065, Japan.}
\altaffiltext{2}{Division of Science, National Astronomical Observatory of Japan, Mitaka, Tokyo 181-8588, Japan.}


\maketitle

\begin{abstract}
We quantify how bar-driven non-circular motions bias Milky-Way gas maps inferred with the kinematic-distance (KD) method. KD reconstructions of H\,\textsc{i} and CO surveys assume circular rotation in an axisymmetric potential, an assumption that is strongly violated in the barred inner Milky Way.
We use high-resolution hydrodynamical simulations of gas flow in an observationally constrained barred Milky Way potential. From a quasi-steady snapshot we generate synthetic longitude--velocity data and apply a standard axisymmetric KD inversion using the circular-speed curve derived from the $m=0$ component of the same potential. To isolate non-circular effects, we remove the near--far ambiguity by selecting, for each gas element, the KD branch closest to its true distance.
We find that the KD method reproduces the gas distribution reasonably well outside the bar-dominated region ($R \gtrsim 5$~kpc), but fails systematically in the bar region ($R \sim 0.5$--3~kpc). There the KD-reconstructed face-on map exhibits anisotropic, quadrant-dependent artifacts, including arc-like overdensities and LOS-elongated low-density cavities. In azimuthally averaged profiles, these anisotropic misassignments translate into net radial mixing: the axisymmetric KD inversion substantially fills in the true bar-induced depression (hereafter, the ``bar gap'') and yields a flatter inner profile. Distance-error maps show coherent structures with $|\Delta d| \sim 1$--2~kpc and relative errors of several tens of percent along the bar and inner ring, coincident with zones where the KD mapping is intrinsically ill-conditioned, quantified by a large geometric sensitivity $|S| \equiv \left|\partial d/\partial V_{\rm LOS}^{\rm circ}\right|$. In these regions the error is well approximated to first order by $\Delta d \simeq S\,\Delta V_{\rm LOS}$, linking KD failures directly to bar-driven streaming velocities.
Our results demonstrate that KD-based gas maps and radial profiles in the barred inner Milky Way require great caution: an axisymmetric KD inversion can partly fill in the true ``bar gap'' and flatten the inner profile.
\end{abstract}


\section{Introduction}
\label{sec:Introduction}

\begin{figure*}
\begin{center}
\includegraphics[width=1.\textwidth]{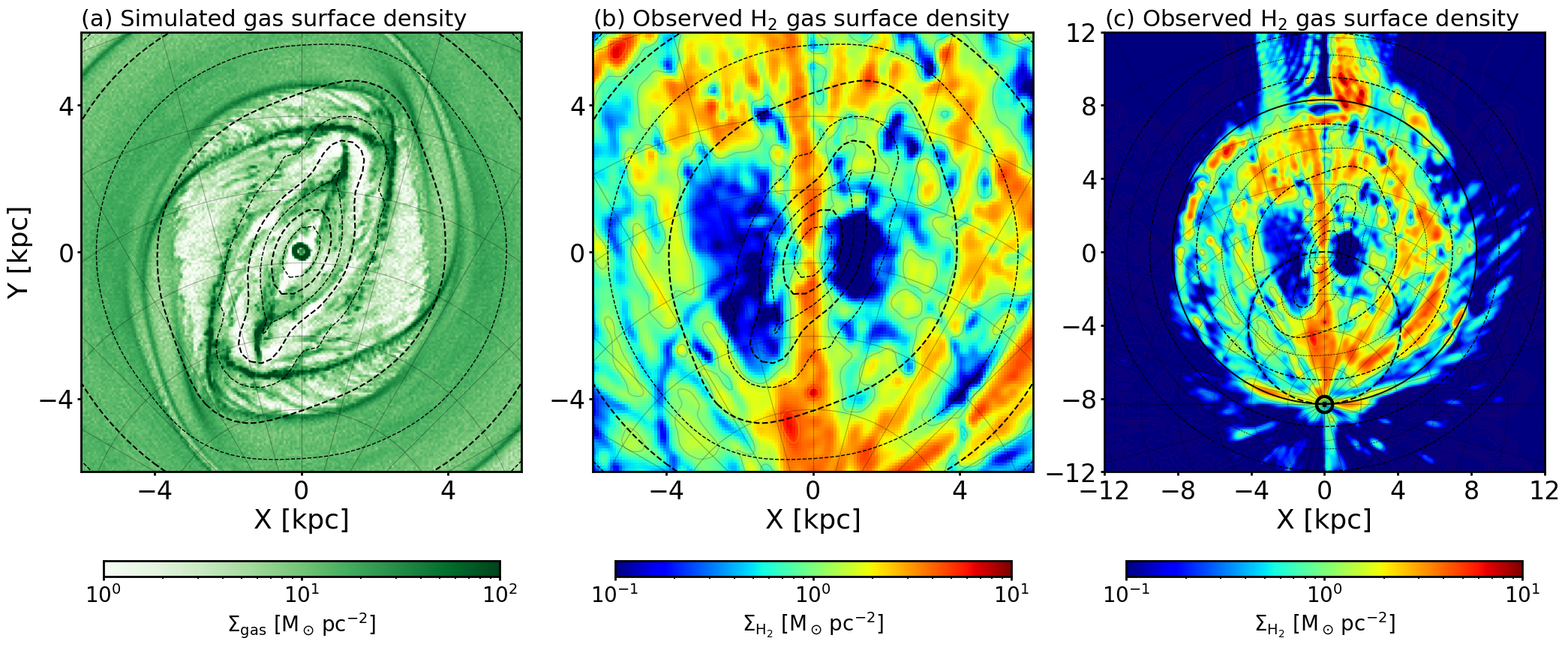}
\end{center}
\caption{
\textbf{(a)} Face-on gas surface-density map from the barred Milky Way hydrodynamical simulation without gas self-gravity presented in \citet{Baba2025b}, overlaid with stellar surface-density contours of the barred stellar component \citep{Portail+2017,Sormani+2022agama}.
\textbf{(b)} Face-on molecular-gas surface-density map, $\Sigma_{\rm H_2}$, reconstructed from the Galactic CO($J$=1--0) survey of \citet{Dame+2001} using the KD method. Each $(\ell,b,v)$ voxel in the data cube is assigned to a Galactocentric position assuming an axisymmetric rotation model, and the near--far ambiguity is resolved following the procedure of \citet{NakanishiSofue2016} (see Appendix).
\textbf{(c)} Same as panel~(b), but shown over a larger radial range. The solid circle marks the Solar circle ($R=R_0$), while the dashed circle indicates the tangent-point locus.
\textbf{Alt Text}: Three face-on maps of gas surface density in the Galactic plane. Panel (a) shows the simulated barred-galaxy gas map with stellar bar density contours. Panel (b) shows the molecular-gas map reconstructed from the Galactic carbon-monoxide J=1-0 survey using an axisymmetric KD method with near-far resolution. Panel (c) shows the same reconstruction over a larger radial range, with circles marking the Solar circle and the tangent-point locus.
}
\label{fig:obs_kd}
\end{figure*}

Mapping the three-dimensional distribution of interstellar gas in the Milky Way is essential for studies of Galactic structure, dynamics, and star formation. Because we observe the Milky Way from within the disk, the gas distribution cannot be obtained by simple projection; it must be reconstructed from sky position and line-of-sight (LOS) velocity. A widely used approach is the kinematic distance (KD) method \citep[e.g.][]{Oort+1958,Burton1971,GeorgelinGeorgelin1976,NakanishiSofue2003}, which assumes that gas follows circular rotation in an axisymmetric potential. Given an adopted rotation curve $V_{\rm c}(R)$, the observed LOS velocity $V_{\rm LOS}(\ell,b)$ can be inverted to estimate a heliocentric distance along each line of sight. KD-based reconstructions applied to H\,\textsc{i} and CO surveys have been used to build three-dimensional gas maps and radial surface-density profiles, and to identify large-scale features such as the molecular ring around $R\sim 4$~kpc \citep[e.g.][]{Bronfman+1988,Dame+2001,NakanishiSofue2006,NakanishiSofue2016,Marasco+2017,Roman-Duval+2009}.

However, the basic KD assumptions are not satisfied in the real Milky Way \citep[][]{Gomez2006,Baba+2009,Hunter+2024}. The Milky Way hosts a strong bar and spiral structure, and its interstellar medium is multi-phase and turbulent. Among these effects, bar-driven streaming motions are expected to be a dominant and coherent source of non-circular velocities in the inner few kiloparsecs \citep[e.g.][]{Binney+1991,Li+2022,Baba2025b}. In this paper we focus on this bar-dominated region, where KD systematics can affect not only individual distances but also global inferences from KD-based maps.

For reference, Figure~\ref{fig:obs_kd}(a) shows an illustrative example of the face-on gas distribution from a hydrodynamical simulation in an observationally constrained barred Milky Way potential \citep[][]{Baba2025b}. In such barred potentials, gas typically flows along offset dust lanes on the leading sides of the stellar bar, concentrates in a CMZ-like nuclear ring, and accumulates near the bar ends and in bar-driven spiral arms. At the same time, the bar tends to deplete gas between $\sim 1$ and 3~kpc, producing a characteristic deficit in this radial range and enhanced concentrations near the bar ends and in the spiral arms \citep[e.g.][]{Athanassoula1992b,EnglmaierGerhard1997,Fux1999,Sormani+2015a,Pettitt+2014,Pettitt+2015,BabaKawata2020a,Sormani+2020,Hunter+2024,Duran-Camacho+2024}.

Against this background, Figure~\ref{fig:obs_kd}(b) presents a KD-reconstructed face-on map of molecular gas inferred from the CO($J$=1--0) survey \citep[][]{Dame+2001}, overlaid with the same barred stellar mass model. If the axisymmetric circular-rotation assumption were adequate, the reconstructed morphology would broadly reflect the barred patterns described above. Instead, the KD map shows spatially asymmetric structures that are difficult to reconcile with a simple barred-gas morphology. In particular, the KD map exhibits ``LOS-elongated low-density cavities'' toward $|\ell|\sim 5^\circ$--$15^\circ$, and the cavity at positive longitudes extends farther along the LOS and covers a larger area. Similar asymmetric cavities have also been reported in previous KD-based reconstructions \citep[][]{NakanishiSofue2016,Sofue2023}.

This mismatch suggests that KD in the barred inner Milky Way is distorted by systematic velocity offsets rather than by random noise alone.
Moreover, KD performance depends strongly on the viewing geometry (i.e. Galactic longitude): along some longitudes the circular-orbit LOS velocity varies only weakly with distance or becomes effectively multivalued, making the inversion intrinsically ill-conditioned \citep[][]{Burton1992,Sofue2011}.
This geometric effect implies that even modest turbulent motions or systematic streaming can produce large distance errors along specific lines of sight. Therefore, to understand KD-based gas maps in the inner Milky Way, one must consider both the non-circular streaming field and the intrinsic geometric sensitivity of the KD mapping.

Recent progress in stellar dynamical modeling provides a timely opportunity to address this problem quantitatively. Made-to-measure (M2M) models have placed strong constraints on the Milky Way bar/bulge mass distribution and pattern speed \citep[][]{Portail+2017,Sormani+2022agama}. Hydrodynamical simulations in such observationally calibrated barred potentials can reproduce the main bar-driven features seen in observed H\,\textsc{i}/CO longitude--velocity diagrams, and thus offer a realistic testbed for KD \citep[][]{Li+2022,Hunter+2024}. In this paper we use high-resolution hydrodynamical simulations of gas flow in an M2M-constrained barred Milky Way potential, apply a standard axisymmetric KD inversion, and compare the KD-reconstructed gas distribution with the known ``true'' distribution in the simulation.
Our goal is not to propose a new distance estimator for individual clouds, but to quantify how bar-driven non-circular motions bias global quantities that are commonly derived from KD reconstructions. In particular, we focus on (i) how KD distorts face-on gas distributions in the bar region; (ii) how the magnitude and sign of KD distance errors relate to the underlying streaming-motion field and to the geometric sensitivity of the method, quantified by $S \equiv \partial d / \partial V_{\rm LOS}$ for a given rotation curve; and (iii) how these distance errors propagate into radial gas surface-density profiles.

The paper is organized as follows. Section~\ref{sec:ModelMethod} describes the barred Milky Way gas-flow simulations, our implementation of the KD inversion, and the diagnostics used to quantify KD failures. Section~\ref{sec:Results} presents the main results, including maps of the true and KD-reconstructed gas distributions, two-dimensional distance-error and KD-sensitivity maps in the Galactic plane, and radial surface-density profiles. Section~\ref{sec:Summary} discusses the implications for KD-based studies of the Milky Way gas distribution and summarizes our conclusions.

\section{Models and Methods}
\label{sec:ModelMethod}

\subsection{Hydrodynamic model in an observationally constrained bar potential}
\label{sec:extpot_model}

We analyze smoothed particle hydrodynamics (SPH) simulations following \citet{Baba2025b}, in which the interstellar gas evolves in an externally imposed Milky Way potential constrained by observations of the inner bar/bulge. The equations of motion are integrated in an inertial Galactocentric frame, and the non-axisymmetric bar is implemented as a rigidly rotating potential. The barred mass distribution follows the M2M model of \citet{Portail+2017}. Outside the bar region, the axisymmetric disk and halo components, as well as the adopted Solar parameters, closely follow the mass model of \citet{Hunter+2024}, with minor adjustments described in \citet{Baba2025b}. The adopted gravitational potential consists of:
\begin{itemize}
  \item a three-dimensional bar+bulge component, represented by the analytic approximation to the M2M model provided by \citet{Sormani+2022agama}, with a pattern speed $\Omega_{\rm b}=37.5~{\rm km~s^{-1}~kpc^{-1}}$;
  \item stellar thin and thick disks and a dark matter halo chosen to reproduce the circular-velocity curve inferred by \citet{Eilers+2019}; and
  \item a nuclear stellar disk and nuclear star cluster as in \citet{Sormani+2020nsd}.
\end{itemize}

Unlike \citet{Hunter+2024}, we do not impose a rigidly rotating stellar spiral perturbation. Since $N$-body simulations indicate that stellar spirals are typically transient, differentialy rotating patterns \citep[e.g.][]{Fujii+2011,Baba+2013}, we omit an imposed spiral potential to isolate the impact of bar-driven non-circular motions; any spiral-like gas features arise purely in response to the bar.

The gas is evolved with the {\tt ASURA} code \citep{Saitoh+2008,SaitohMakino2013}, including radiative cooling and heating, stochastic star formation, and feedback from supernovae and H\,\textsc{ii} regions following \citet{Baba2025b}.
In contrast to \citet{Baba2025b}, we include the self-gravity of the gas, while keeping all other numerical and physical ingredients unchanged.
The external potential is fixed in the bar's rotating frame, so the bar amplitude, length, and orientation remain constant, and the gas responds dynamically to this time-independent background potential plus its self-gravity.

We analyze a representative quasi-steady snapshot in which the gas distribution and kinematics exhibit a CMZ-like nuclear ring and offset dust-lane streams broadly similar to those inferred in the inner Milky Way.

For the KD inversion (see Section~\ref{sec:KDmethod}) we adopt the circular-velocity curve $V_{\rm c}(R)$ associated with the axisymmetric ($m=0$) component of the same gravitational potential \citep[see Fig.~1 in][]{Baba2025b}.
In practice, we expand the full gravitational potential $\Phi(R,\phi,z)$ in a Fourier series in the azimuthal angle,
\begin{equation}
    \Phi(R,\phi,z)
    = \sum_{m=-\infty}^{+\infty} \Phi_m(R,z)\,e^{i m \phi},
\end{equation}
and identify the $m=0$ term, $\Phi_0(R,z)$, as the axisymmetric component.
We then define the corresponding circular velocity curve as
\begin{equation}
    V_{\rm c}^2(R) \equiv
    R\,\left.\frac{\partial \Phi_0}{\partial R}\right|_{z=0},
    \label{eq:vc}
\end{equation}
which we take as the ``true'' rotation curve underlying the gas flow.
This $V_{\rm c}(R)$ is used consistently in the KD model described in Section~\ref{sec:KDmethod}.

For comparison with common observational practice, we also repeat the KD reconstruction using a terminal-velocity-based rotation curve in Section~\ref{sec:kd_obs_tvc}. This alternative choice does not change our main conclusions.

\subsection{Kinematic distance method}
\label{sec:KDmethod}

\begin{figure*}
\begin{center}
\includegraphics[width=1.0\textwidth]{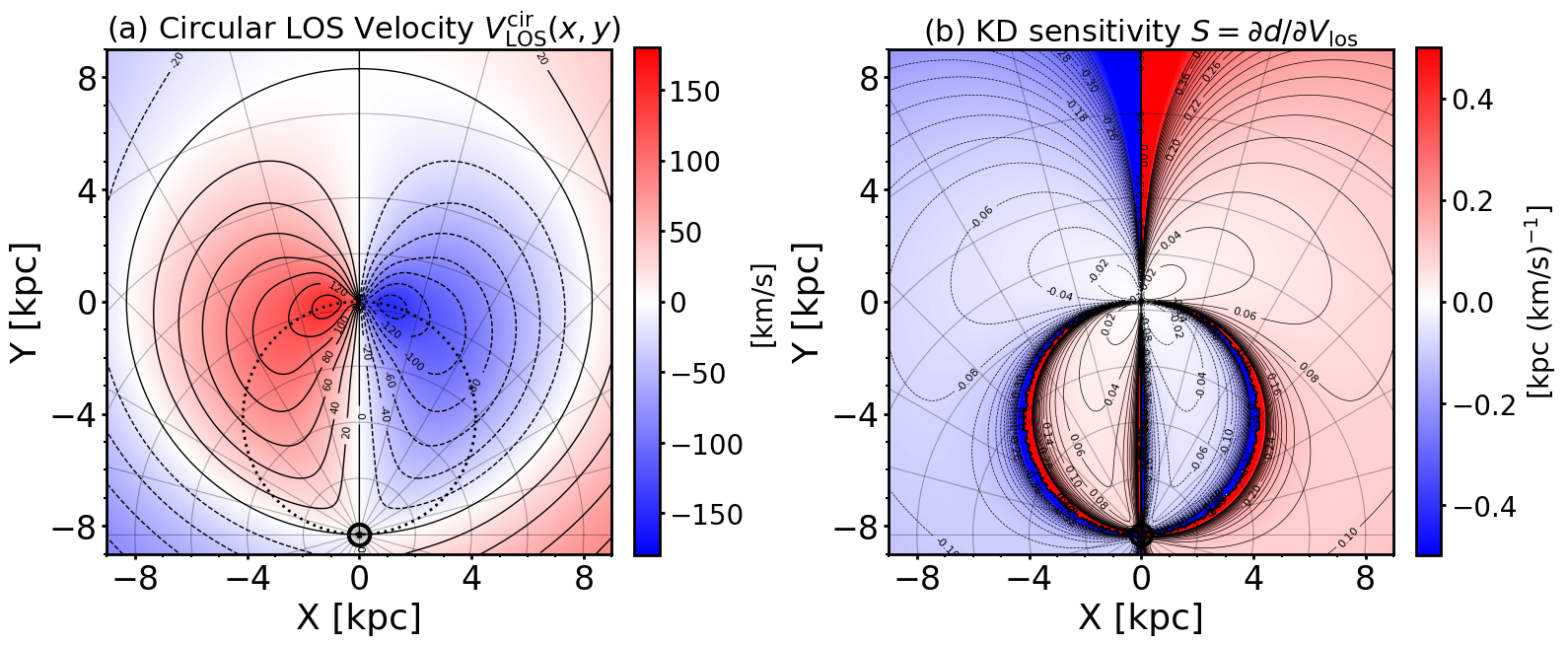}
\end{center}
\caption{
\textbf{(a)} Model circular LOS velocity field, $V_{\rm LOS}^{\rm circ}(X,Y)$, computed from the adopted axisymmetric rotation curve. The dashed circle indicates the tangent-point locus.
\textbf{(b)} KD-sensitivity map, $S(X,Y)$, defined from the response of the circular-orbit distance to perturbations in LOS velocity.
\textbf{Alt Text}: Two face-on maps in the Galactic plane illustrating the axisymmetric KD model. Panel (a) shows the circular line-of-sight velocity field predicted from the adopted rotation curve, with a dashed circle marking the tangent-point locus and the Sun marked by a symbol. Panel (b) shows the KD sensitivity $S$, defined as the change in inferred distance per change in line-of-sight velocity, with the same tangent-point circle overplotted.
}
\label{fig:kd_method}
\end{figure*}

\begin{figure}
\begin{center}
\includegraphics[width=0.45\textwidth]{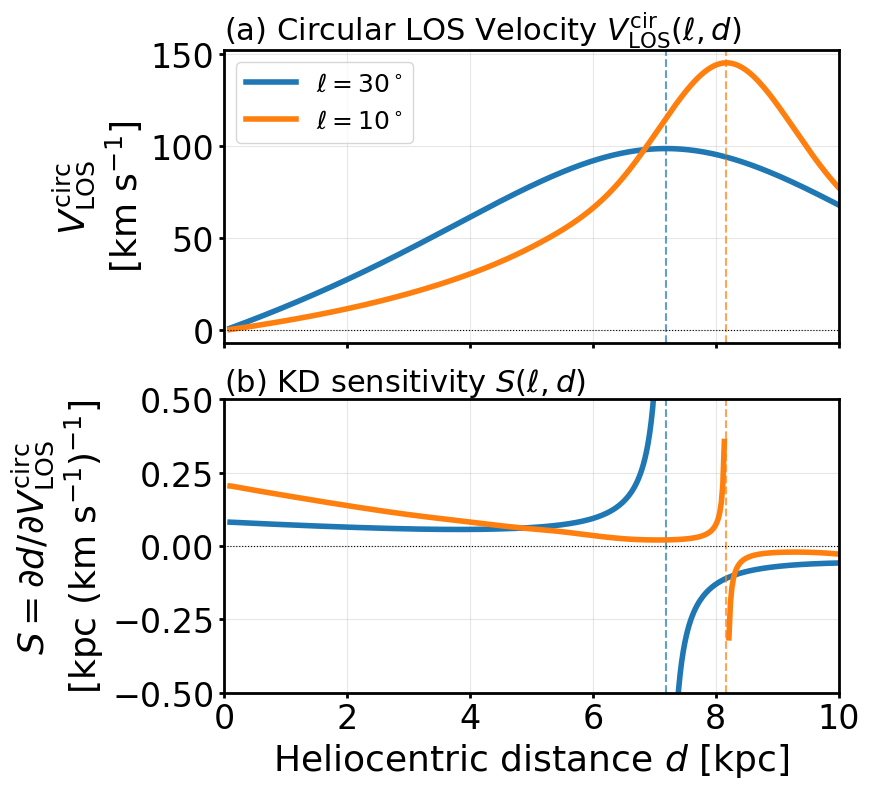}
\end{center}
\caption{
Distance dependence of \textbf{(a)} the circular LOS velocity $V_{\rm LOS}^{\rm circ}(\ell,d)$. and \textbf{(b)} the KD sensitivity $S(\ell,d)$ along the selected Galactic longitudes in the Galactic plane.
Dashed vertical lines mark the tangent-point distance along each line of sight, defined as the location where $|V_{\rm LOS}^{\rm circ}(d)|$ reaches its maximum.
\textbf{Alt Text}: Two line plots show the axisymmetric KD model versus heliocentric distance for selected Galactic longitudes in the Galactic plane. Panel (a) shows the circular line-of-sight velocity as a function of distance. Panel (b) shows the KD sensitivity $S$ as a function of distance. Dashed vertical lines mark the tangent-point distance where the absolute circular line-of-sight velocity reaches its maximum.
}
\label{fig:vlos_sens_model}
\end{figure}

For the chosen snapshot we place a virtual observer at a Sun--like position in the Galactic plane.
We adopt a Galactocentric distance $R_0 = 8.3$~kpc and a circular speed $V_0 = V_{\rm c}(R_0) = 230~\rm km~s^{-1}$, and locate the observer at Galactocentric Cartesian coordinates $(X,Y,Z) = (0,-R_0,0)$ in the simulation frame.
For each SPH particle we compute the heliocentric distance $d$ and Galactic longitude and latitude $(\ell,b)$ from its Cartesian coordinates.
We then calculate the line--of--sight (LOS) velocity relative to the local standard of rest (LSR) as
\begin{equation}
    V_{\rm LOS} =
    \bigl[ \mathbf{v} - \mathbf{V}_{\rm obs} \bigr] \cdot
    \hat{\mathbf{n}}_{\rm LOS},
\end{equation}
where $\mathbf{v}$ is the gas velocity, $\mathbf{V}_{\rm obs}$ is the velocity of the observer (taken to be $(0,V_0,0)$ in Galactocentric Cartesian coordinates), and $\hat{\mathbf{n}}_{\rm LOS}$ is the unit vector from the observer to the SPH particle.
By binning particles in $(\ell,V_{\rm LOS})$ we can construct synthetic longitude--velocity ($\ell$--$v$) diagrams, although in this paper we mainly use the individual $(\ell,b,V_{\rm LOS})$ values for the KD inversion.

To mimic the standard KD method, we assume axisymmetric circular rotation described by $V_{\rm c}(R)$ and neglect any non--circular motions.
For a given $(\ell,b)$ and heliocentric distance $d$, the Galactocentric radius is
\begin{equation}
    R(d,\ell,b) =
    \sqrt{R_0^2 + d^2\cos^2 b - 2 R_0 d \cos b \cos\ell},
\end{equation}
and under pure circular rotation the expected LOS velocity is
\begin{equation}
    V_{\rm LOS}^{\rm circ}(d,\ell,b) =
    \Biggl[
      \frac{R_0 \sin\ell}{R(d,\ell,b)}\,
      V_{\rm c}\bigl(R(d,\ell,b)\bigr)
      - V_0 \sin\ell
    \Biggr] \cos b.
    \label{eq:vlos_circ}
\end{equation}
Figure~\ref{fig:kd_method}(a) shows the corresponding model LOS velocity field, $V_{\rm LOS}^{\rm circ}(X,Y)$, in the Galactic plane. 

In practice, instead of solving $V_{\rm LOS}^{\rm circ}(d,\ell,b) = V_{\rm LOS}$ analytically for each particle, we perform a numerical inversion along the corresponding line of sight.
We sample heliocentric distance on a regular grid $s \in (0.01,40)$~kpc and evaluate the circular--rotation model $V_{\rm LOS}^{\rm circ}(s,\ell,b)$ at each grid point.
We then model the scatter between the observed and model LOS velocities with a fixed one--dimensional turbulent dispersion $\sigma_{\rm turb}$ and define a Gaussian likelihood
\begin{equation}
    \mathcal{L}(s) \;\propto\;
    \exp\!\left[
      -\frac{1}{2}
      \left(
        \frac{V_{\rm LOS} - V_{\rm LOS}^{\rm circ}(s,\ell,b)}
             {\sigma_{\rm turb}}
      \right)^2
    \right].
\end{equation}
Assuming a flat prior in distance, the posterior probability $p(s \mid V_{\rm LOS})$ is proportional to $\mathcal{L}(s)$, and we define the kinematic distance as the maximum--a--posteriori (MAP) estimate
\begin{equation}
    s_{\rm MAP} \equiv \arg\max_s \mathcal{L}(s),
\end{equation}
which in this case coincides with the maximum--likelihood estimate.
In our fiducial analysis we adopt $\sigma_{\rm turb} = 7~{\rm km\,s^{-1}}$.

Figure~\ref{fig:vlos_sens_model}(a) illustrates the distance dependence of $V_{\rm LOS}^{\rm circ}(\ell,d)$ along selected longitudes $\ell = 10^\circ$ and $30^\circ$ in the midplane ($b=0^\circ$), and marks the tangent-point distances at which $|V_{\rm LOS}^{\rm circ}|$ reaches an extremum. In the inner Milky Way ($|\ell|<90^\circ$), the circular-rotation relation is generally double-valued:  a given $V_{\rm LOS}^{\rm circ}$ corresponds to two heliocentric distances (the classical near--far ambiguity), except at the tangent point.

In this work, however, we aim to isolate only the impact of non--circular motions on the KD method.
We therefore make use of the fact that in the simulation the true distance $d_{\rm true}$ of each SPH particle is known.
Whenever two circular--rotation solutions $d_{\rm kin}^{\rm (near)}$ and $d_{\rm kin}^{\rm (far)}$ exist for a given $(\ell,b,V_{\rm LOS})$, we assign
\begin{equation}
    d_{\rm kin} =
    \arg\min_{d \in \{d_{\rm kin}^{\rm (near)},\,d_{\rm kin}^{\rm (far)}\}}
      \bigl| d - d_{\rm true} \bigr|,
\end{equation}
i.e.\ we select the branch whose circular--rotation distance lies closest to the true distance of that SPH particle.
If no distinct near/far peaks are found, we simply adopt $d_{\rm kin} = s_{\rm MAP}$.
This procedure removes the classical near--far ambiguity while preserving the basic structure of the KD inversion, allowing us to focus on the biases produced specifically by bar--induced non--circular motions.

\subsection{Geometric sensitivity of kinematic distance}
\label{sec:KDSensitivity}

In addition to the simulation-based diagnostic, we estimate the purely geometric sensitivity of the KD method by computing the response of the circular-orbit distance to changes in LOS velocity, in a spirit similar to analytic treatments of KD geometry \citep[][]{Sofue2011}. Along a given line of sight $(\ell,b)$, the model LOS velocity under circular rotation is given by equation~(\ref{eq:vlos_circ}). For a grid of points in the Galactic plane we evaluate $V_{\rm LOS}^{\rm circ}(d,\ell,b)$ and estimate the derivative $\partial V_{\rm LOS}^{\rm circ} / \partial d$ numerically. Whenever this derivative is significantly different from zero, we define the ``KD sensitivity''
\begin{equation}
 S(X,Y) \equiv \frac{\partial d}{\partial V_{\rm LOS}^{\rm circ}}
    = \left(\frac{\partial V_{\rm LOS}^{\rm circ}}{\partial d}\right)^{-1},
\end{equation}
which measures how a small velocity perturbation translates into a distance error along the corresponding line of sight. 
This quantity is closely related to the classical ``velocity crowding'' parameter $\partial V_{\rm LOS}^{\rm circ}/\partial d$ introduced by \citet{Burton1971} and discussed in detail by \citet{Burton1992}.

Figure~\ref{fig:kd_method}(b) displays the resulting KD-sensitivity map $S(X,Y)$. As expected, the sensitivity diverges along directions where the circular-orbit LOS velocity is either identically zero or reaches an extremum as a function of distance. Along the Galactic center and anticenter lines ($\ell \approx 0^\circ$ and $180^\circ$), the circular-orbit LOS velocity $V_{\rm LOS}^{\rm cir}$ vanishes for all distances, so $V_{\rm LOS}^{\rm circ}$ carries no information on distance and $S(X,Y)$ formally becomes infinite. Likewise, along the tangent point circle the LOS velocity attains a maximum or minimum with respect to $d$, i.e. $\partial V_{\rm LOS}^{\rm circ}/\partial d = 0$ at the tangent point, and a given velocity corresponds to two distinct distances (the usual near- and far-side solutions). In these regions the KD method is intrinsically ill-conditioned: even small random or systematic perturbations in velocity map into very large, and often multivalued, distance errors \citep[][]{Sofue2011}.

The same behavior is evident in the one-dimensional view shown in Figure~\ref{fig:vlos_sens_model}(b), which plots $S(\ell,d)$ along selected longitudes $\ell = 10^\circ$ and $30^\circ$ in the midplane ($b=0^\circ$). The sensitivity exhibits sharp excursions and sign changes in the vicinity of the tangent point, where $\partial V_{\rm LOS}^{\rm circ}/\partial d \rightarrow 0$, and remains small in magnitude where $V_{\rm LOS}^{\rm circ}(\ell,d)$ varies monotonically and steeply with distance. Therefore, even in a purely circular model, lines of sight that pass close to a tangent-point extremum are highly susceptible to both random (turbulent) and systematic (streaming) velocity offsets.

Away from these singular loci, the sign of $S(X,Y)$ follows a simple pattern set by the KD geometry. Outside the tangent-point circle, $S<0$ at $\ell>0^\circ$ and $S>0$ at $\ell<0^\circ$, whereas inside the tangent-point circle the sign reverses, with $S>0$ for $\ell>0^\circ$ and $S<0$ for $\ell<0^\circ$. This behavior reflects whether the LOS circular velocity $V_{\rm LOS}^{\rm cir}$ increases or decreases with distance along a given line of sight, and thus whether a positive velocity perturbation tends to push the inferred KD distance farther from, or closer to, the observer (see Figures~\ref{fig:kd_method}(a) and \ref{fig:vlos_sens_model}(a)).

\section{Results}
\label{sec:Results}

\subsection{Face-on gas distribution and streaming motions}
\label{subsec:maps_streaming}

\begin{figure*}
\begin{center}
\includegraphics[width=1.0\textwidth]{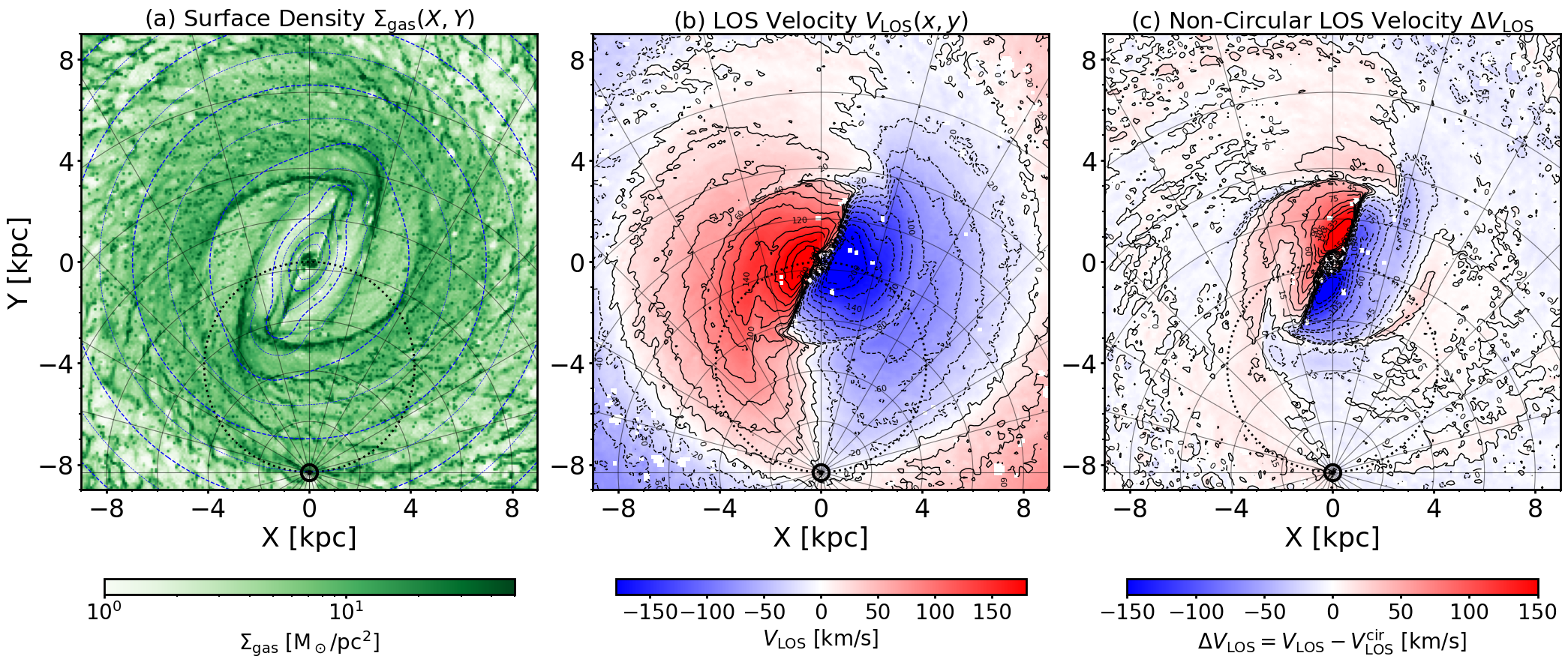}
\end{center}
\caption{
Face-on gas distribution, LOS velocity fields, and non-circular component of the LOS velocity in the fiducial barred Milky Way simulation.
All panels show the Galactic plane in Cartesian coordinates $(X,Y)$, with the Galactic center at the origin and the Sun at $(0,-R_0)$.
The dotted circle shows the tangent-point locations (i.e. terminal velocity circle).
\textbf{(a)} True gas surface density, $\Sigma_{\rm gas}(X,Y)$, obtained by vertically integrating the gas density.
The bar appears as an elongated high-density feature tilted by $25^\circ$ clockwise from the $Y$-axis, accompanied by a CMZ-like nuclear ring and bar-driven spiral arms.
\textbf{(b)} Mass-weighted mean LOS velocity field, $V_{\rm LOS}^{\rm true}(X,Y)$.
Contours highlight strong streaming motions along the bar and spiral features.
\textbf{(c)} Non-circular component of the LOS velocity,
$\Delta V_{\rm LOS}(X,Y) = V_{\rm LOS}^{\rm true} - V_{\rm LOS}^{\rm circ}$,
where $V_{\rm LOS}^{\rm circ}$ is the LOS velocity field predicted by the axisymmetric circular-rotation model used in the KD method.
\textbf{Alt Text}: Three face-on maps in the Galactic plane from the fiducial hydrodynamic simulation, in Cartesian $X$ and $Y$ with the Galactic center at the origin and the Sun at $(0, -R_0)$. A dotted circle marks the tangent-point locus. Panel (a) shows the true gas surface density with an elongated bar and a central ring. Panel (b) shows the mean line-of-sight velocity field. Panel (c) shows the non-circular line-of-sight velocity component, defined as the true field minus the axisymmetric circular-rotation prediction.
}
\label{fig:truemap_v2}
\end{figure*}

Figure~\ref{fig:truemap_v2}(a) shows the true gas surface density, $\Sigma_{\rm gas}(X,Y)$, obtained by vertically integrating the gas density in the snapshot and binning it on a regular Cartesian grid. 
The bar appears as an elongated high--density feature whose major axis is tilted by $25^\circ$ clockwise from the $Y$--axis (i.e., from the Sun--center line) and is accompanied by a CMZ--like nuclear ring and prominent bar--driven spiral arms outside the bar region.
Because no stellar spiral arms are included in the external potential, spiral structure in this model is induced primarily by the bar, consistent with earlier hydrodynamical simulations of gas in barred potentials \citep[e.g.][]{Schwarz1981,CombesGerin1985,Athanassoula1992b,EnglmaierGerhard1997,Bissantz+2003}.  
The Sun is marked by the $\odot$ symbol at $(X,Y) = (0,-R_0)$. 

Figure~\ref{fig:truemap_v2}(b) displays the mass--weighted mean LOS velocity field, $V_{\rm LOS}(X,Y)$, as seen by the observer at $(0,-R_0,0)$.
To first order, the pattern resembles the circular--rotation field in Figure~\ref{fig:kd_method}(a): gas at positive longitudes is predominantly redshifted ($V_{\rm LOS}>0$), while gas at negative longitudes is predominantly blueshifted ($V_{\rm LOS}<0$).
However, because the Sun views the bar from an angle of about $25^\circ$ from its major axis, the actual velocity field is not symmetric about $\ell = 0^\circ$.
In particular, there are substantial regions with $V_{\rm LOS}<0$ at $\ell>0^\circ$ between the Sun and the bar, and regions with $V_{\rm LOS}>0$ at $\ell<0^\circ$ near the bar major axis.
The iso--velocity contours are strongly twisted and kinked along the bar--driven spiral arms, revealing large non--circular streaming motions that cannot be captured by a simple axisymmetric circular--rotation model.
In the present simulation we do not include any explicit stellar spiral--arm potential, so in the real Milky Way additional non--circular motions driven by stellar spirals would be expected to distort the iso--velocity contours even further.

To make these deviations explicit, Figure~\ref{fig:truemap_v2}(c) plots the non--circular LOS component,
\begin{equation}
    \Delta V_{\rm LOS}(X,Y)
    \equiv V_{\rm LOS}^{\rm true}(X,Y)
         - V_{\rm LOS}^{\rm circ}(X,Y),
    \label{eq:vlos_diff}
\end{equation}
where $V_{\rm LOS}^{\rm circ}(X,Y)$ is the circular--orbit LOS velocity field shown in Figure~\ref{fig:kd_method}(a). 
Coherent regions with $|\Delta V_{\rm LOS}| \gtrsim 50$--$100~{\rm km\,s^{-1}}$ are concentrated in the bar--dominated inner few kiloparsecs, especially along the bar major axis.
These are precisely the regions in which the circular--orbit assumption underlying the KD method is expected to break down most severely.

\subsection{KD-reconstructed gas maps}
\label{subsec:true_vs_kd_maps}

\begin{figure*}
\begin{center}
\includegraphics[width=1.0\textwidth]{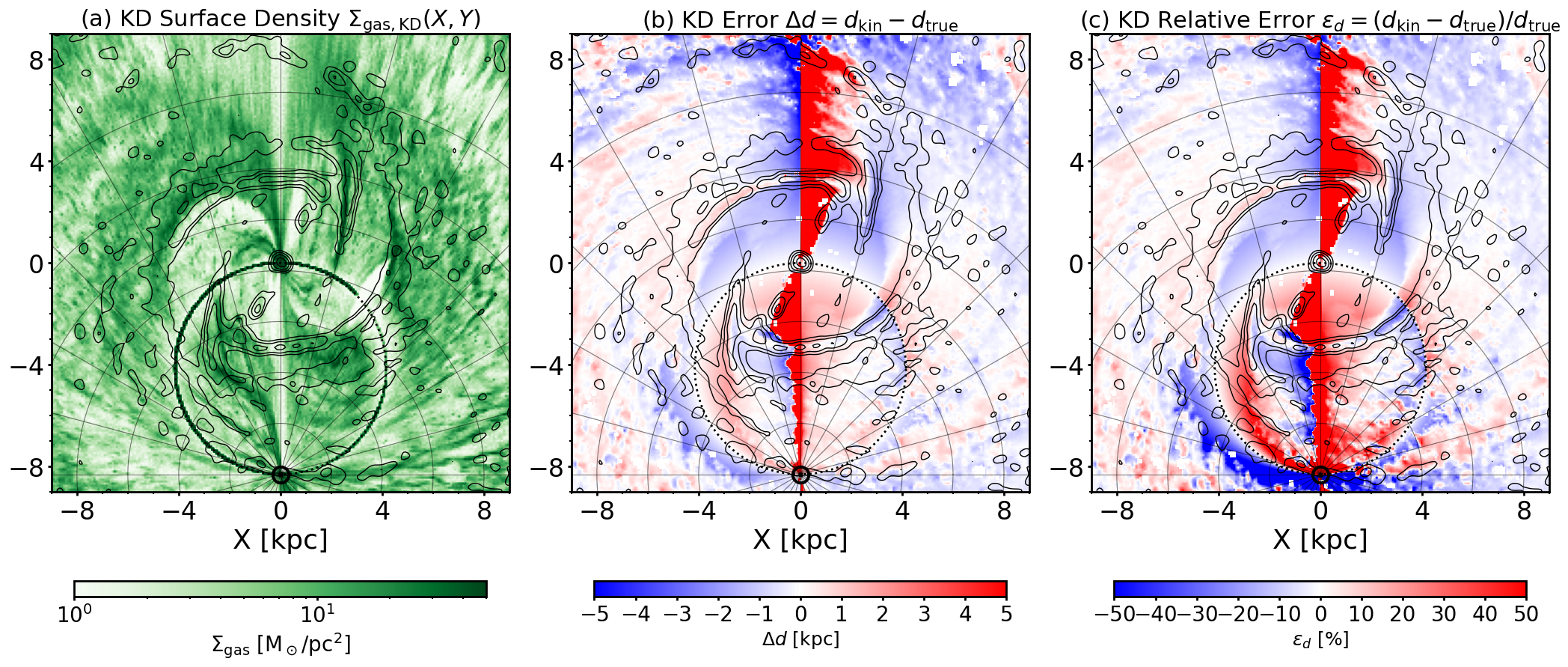}
\end{center}
\caption{
\textbf{(a)} KD-based surface-density map, $\Sigma_{\rm gas,KD}(X,Y)$, obtained by inverting synthetic longitude\UTF{2013}velocity data under the assumption of purely circular rotation and, for each SPH particle, selecting the KD solution (near or far) closest to its true distance. The same color scale as in Figure~\ref{fig:truemap_v2}(a) is used, and black contours show the true surface density from Figure~\ref{fig:truemap_v2}(a).
\textbf{(b)} Mean KD error, $\Delta d(X,Y)$.
\textbf{(c)} Mean KD relative error, $\epsilon_d(X,Y)$.
\textbf{Alt Text}: Three face-on maps in the Galactic plane comparing KD reconstruction to the true simulation. Panel (a) shows the KD surface-density map, with black contours marking the true surface density. Panel (b) shows the mean distance error in kiloparsecs. Panel (c) shows the mean relative distance error, highlighting coherent, bar-related regions of over- and under-estimated distances.
}
\label{fig:kd_map_v2}
\end{figure*}

We apply the axisymmetric KD inversion described in Section~\ref{sec:ModelMethod} to the synthetic $(\ell,v)$ data constructed from our fiducial hydrodynamic simulation.
Figure~\ref{fig:kd_map_v2}(a) shows the KD-reconstructed face-on surface-density map, $\Sigma_{\rm gas,KD}(X,Y)$, for our fiducial model. The black contours reproduce the true surface density, $\Sigma_{\rm gas,true}(X,Y)$, measured directly from the simulation snapshot. Overall, the KD reconstruction is broadly reliable outside the bar-dominated region ($R \gtrsim 5$~kpc), but it exhibits clear, spatially coherent failures in the bar-dominated inner Milky Way ($R \lesssim 4$~kpc).

Beyond the bar region ($R \gtrsim 5$~kpc), the KD map recovers both the overall surface-density level and the main spiral features reasonably well. This behavior is expected because the non-circular component of the LOS velocity, $|\Delta V_{\rm LOS}|$, is relatively small at these radii (Figure~\ref{fig:truemap_v2}c), so that the axisymmetric circular-rotation model provides a good approximation for the inversion.
We note, however, that our model does not include a stellar spiral potential, and thus may underestimate non-circular motions outside the bar region; the KD performance at $R \gtrsim 5$~kpc could be worse in a model with stronger spiral-driven streaming \citep[e.g.][]{Gomez2006,Baba+2009,Hunter+2024}.

By contrast, in the inner Milky Way ($R \lesssim 4$~kpc) the KD-reconstructed map shows pronounced, systematic distortions associated with bar-driven streaming motions. First, the map develops arc-like overdense and underdense ridges that are not present in the true distribution, broadly aligned with the bar-induced flow pattern. Second, it produces ``LOS-elongated low-density cavities'', i.e., underdense regions stretched along the line of sight, which are difficult to reconcile with a simple barred-gas morphology. Third, compact inner features such as the CMZ-like nuclear ring and the inner ring are weakened and azimuthally smeared, and the apparent radius and shape of the ``molecular ring'' can be shifted relative to the true high-density ring. These characteristic artifacts are reminiscent of distortions reported in KD-based reconstructions of observed Galactic gas surveys \citep[e.g.][]{Sofue2023}.

To quantify where the classical axisymmetric KD method is reliable, we compare, for each SPH particle, its true heliocentric distance $d_{\rm true}$ with the kinematic distance $d_{\rm kin}$. We then construct mass-weighted maps of the mean distance error
\begin{equation}
    \Delta d(X,Y) \equiv
    \left\langle d_{\rm kin} - d_{\rm true} \right\rangle,
    \label{eq:kd_error}
\end{equation}
and of the mean relative error
\begin{equation}
    \epsilon_d(X,Y) \equiv
    \left\langle
      \frac{d_{\rm kin} - d_{\rm true}}{d_{\rm true}}
    \right\rangle,
\end{equation}
where the averages are taken over all SPH particles in each $(X,Y)$ cell, weighted by their gas mass.

Figure~\ref{fig:kd_map_v2}(b) shows the mean distance-error map, $\Delta d(X,Y)$, masked where the true gas surface density is very low. The map reveals spatially coherent regions where KD systematically overestimates ($\Delta d>0$) or underestimates ($\Delta d<0$) distances. These patches closely follow the bar, dust lanes, and inner ring, demonstrating that bar-driven streaming motions (Figure~\ref{fig:truemap_v2}c) translate into systematic KD biases of order 1--2~kpc, and locally up to a few kiloparsecs. A notable geometric signature is that the error often flips sign across the terminal-velocity circle (dotted), and the most extreme errors occur toward $\ell \approx 0^\circ$. In contrast, outside the bar-dominated region ($R \gtrsim 5$~kpc in our present setup), $\Delta d$ is typically small and shows little large-scale coherence.

Figure~\ref{fig:kd_map_v2}(c) shows the corresponding map of the mean relative error, $\epsilon_d(X,Y)$.
Its overall morphology broadly parallels that of $\Delta d(X,Y)$, again highlighting the bar, dust lanes, and inner ring as the regions where the KD method performs worst.
However, because the errors are now normalized by $d_{\rm true}$, the relative error map emphasizes sightlines where the true distance is small: even modest absolute errors translate into large fractional biases.
In much of the barred inner disk, $|\epsilon_d|$ reaches 10-20\%, whereas outside the bar region ($R \gtrsim 5$~kpc) both the absolute and relative KD errors remain comparatively small. 

We will return to both features in Section~\ref{subsec:distance_errors}, where we show that the sign flip and the large errors near $\ell\approx 0^\circ$ are primarily controlled by the KD sensitivity $S$ through $\Delta d \simeq S\,\Delta V_{\rm LOS}$.

\subsection{Geometric sensitivity-based prediction}
\label{subsec:distance_errors}

\begin{figure*}
\begin{center}
\includegraphics[width=1.\textwidth]{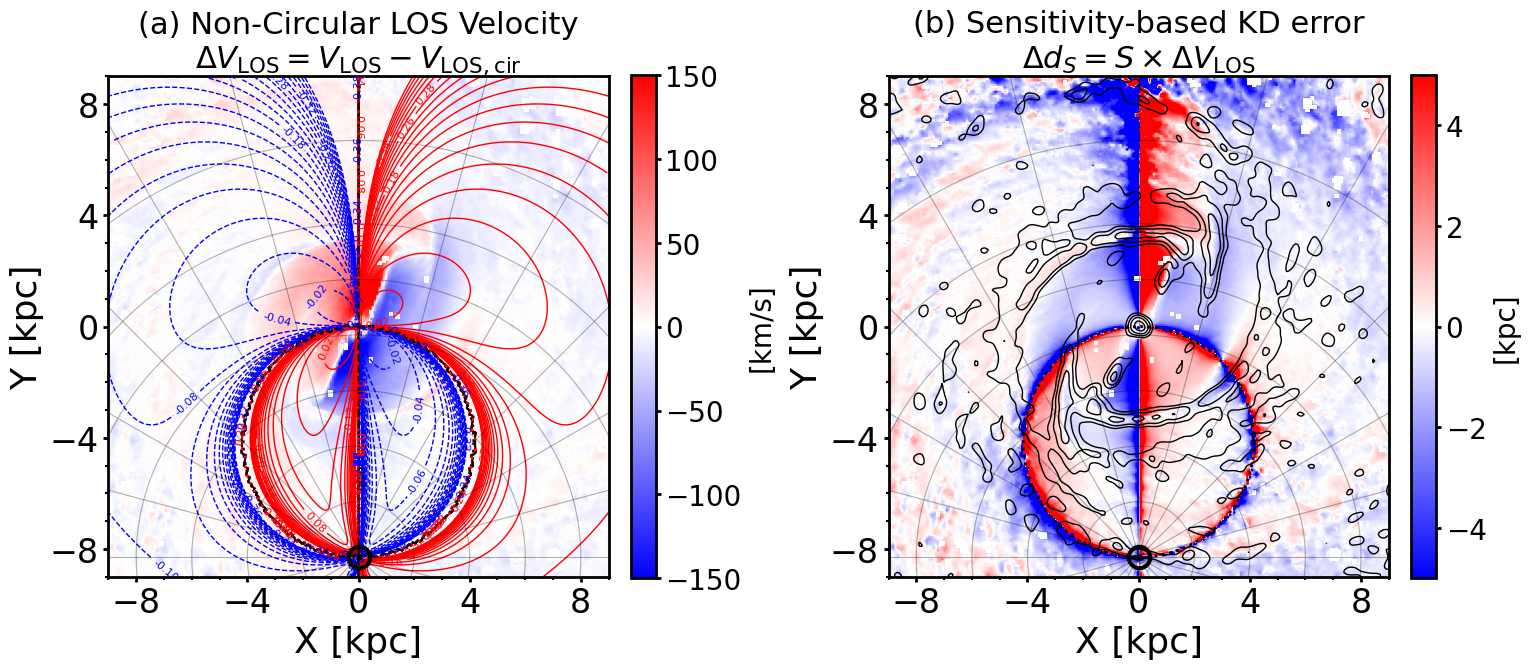}
\end{center}
\caption{
\textbf{(a)} Non-circular LOS-velocity field, $\Delta V_{\rm LOS}(X,Y)$, with KD-sensitivity contours overplotted (red solid for $S>0$ and blue dashed for $S<0$).
\textbf{(b)} First-order, sensitivity-based prediction of the KD error,
$\Delta d_{S}(X,Y) = S(X,Y)\,\Delta V_{\rm LOS}(X,Y)$, constructed by combining the KD sensitivity with the simulated non-circular streaming.
\textbf{Alt Text}: Two face-on maps in the Galactic plane linking non-circular motions to KD errors. Panel (a) shows the non-circular line-of-sight velocity field, with contours indicating the sign of the KD sensitivity $S$. Panel (b) shows the first-order predicted distance error from sensitivity times non-circular velocity, highlighting where the KD method is expected to over- or under-estimate distances.
}
\label{fig:S_map_v2}
\end{figure*}

Comparing the KD-sensitivity map $S(X,Y)$ (Figure~\ref{fig:kd_method}) with the simulation-based error maps $\Delta d(X,Y)$ and $\epsilon_d(X,Y)$, we find that regions with large $|S(X,Y)|$ broadly coincide with those where the mean distance bias is largest, namely along the bar. To interpret the pattern of the actual bar-induced distance errors, we make a simple first-order, sensitivity-based prediction that combines the KD sensitivity with the non-circular streaming. For each SPH particle, we take the LOS-velocity residual $\Delta V_{\rm LOS}$ and evaluate the KD sensitivity at its projected position, $S(X,Y)$. We then construct a sensitivity-based prediction for the corresponding distance error,
\begin{equation}
    \Delta d_{S}(X,Y) \equiv S(X,Y)\,\times\,\Delta V_{\rm LOS}(X,Y),
\end{equation}
which propagates the non-circular velocity perturbations through the KD mapping between distance and LOS velocity.

Figure~\ref{fig:S_map_v2}(a) shows the non-circular LOS-velocity field $\Delta V_{\rm LOS}(X,Y)$ as a color map, with KD-sensitivity contours overplotted. Red solid contours mark regions with $S>0$, while blue dashed contours mark regions with $S<0$. The color map uses red for $\Delta V_{\rm LOS}>0$ and blue for $\Delta V_{\rm LOS}<0$. In this representation, the sign of the expected KD distance error follows directly from the combination of the two signs: where $\Delta V_{\rm LOS}$ and $S$ have the same sign, the first-order prediction gives $\Delta d_{S}>0$ (gas placed too far away), whereas where their signs differ, $\Delta d_{S}<0$ (gas placed too close). 

Figure~\ref{fig:S_map_v2}(b) shows the resulting sensitivity-based prediction map $\Delta d_{S}(X,Y)$. Remarkably, the predicted pattern $\Delta d_{S}(X,Y)$ closely reproduces the main features of the true KD error map $\Delta d(X,Y)$ in Figure~\ref{fig:kd_map_v2}(b), especially along the bar. This agreement demonstrates that the large, coherent biases in the KD-assigned distances can be understood, to leading order, as the product of the KD geometry (encoded in $S(X,Y)$) and the bar-driven streaming velocities (encoded in $\Delta V_{\rm LOS}$). Deviations between $\Delta d_{S}$ and $\Delta d$ mainly occur near the tangent point circle and other multivalued regions, where the KD solution is intrinsically non-linear and the simple first-order approximation breaks down.

\subsection{Radial surface-density profiles}
\label{subsec:radial_profiles}

Barred-potential hydrodynamic simulations, including our fiducial run, naturally produce a CMZ, a molecular ring at a few kiloparsecs, and a relative deficit of molecular gas at intermediate radii ($R\sim 0.5$--3~kpc). Hereafter, we call this bar-related molecular-gas deficit the ``bar gap''.
In the inner Milky Way, however, KD estimates are particularly sensitive to non-circular motions and to the unfavorable KD geometry, so radial surface-density profiles inferred from KD-based face-on maps need not match the true underlying profile.

To compress the two-dimensional maps into compact diagnostics, we compute radial profiles from both the true and KD-reconstructed surface-density maps.
For a given surface-density field $\Sigma_{\rm gas}(X,Y)$, we define the azimuthally averaged profile $\langle \Sigma_{\rm gas}(R)\rangle$.
Because KD assigns gas to incorrect distances in a highly anisotropic, quadrant-dependent manner in the bar region (Sections~\ref{subsec:true_vs_kd_maps} and \ref{subsec:distance_errors}), the KD face-on map develops LOS-elongated cavities/ridges and, more generally, misplaces gas across a wide range of Galactocentric radii.
Consequently, the azimuthally averaged profile $\langle\Sigma_{\rm gas}(R)\rangle$ reflects this net radial redistribution after mixing coherent overdense and underdense sectors, and its bias need not have the same sign as local deficits seen in specific regions of the two-dimensional map.

\begin{figure}
\begin{center}
\includegraphics[width=0.45\textwidth]{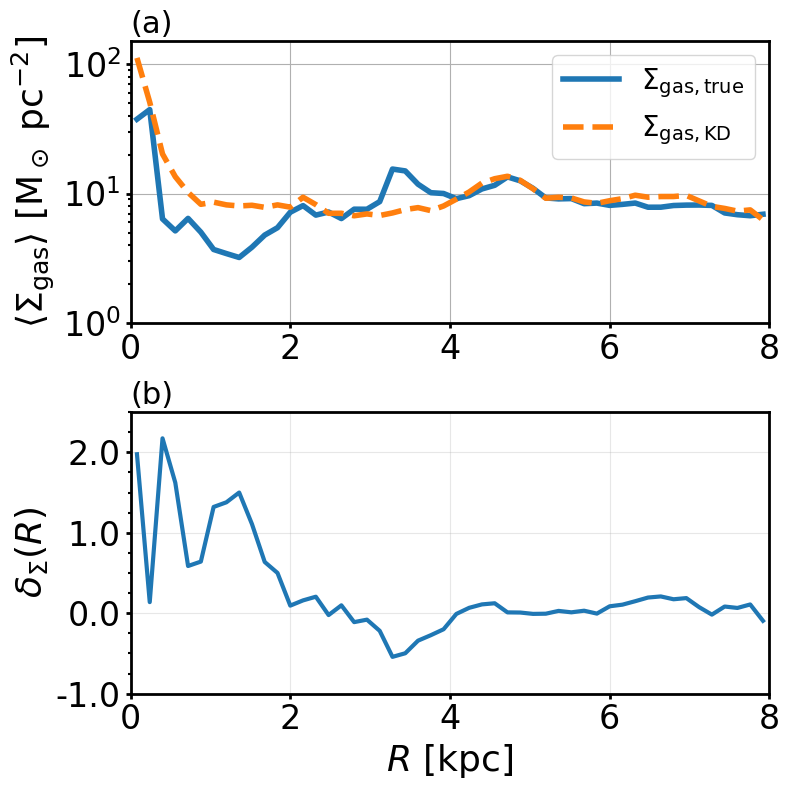}
\end{center}
\caption{
Azimuthally averaged radial surface-density profiles in the Galactic plane. \textbf{(a)} Comparison between the profile measured from the true face-on gas distribution in the simulation, $\langle \Sigma_{\rm gas,true}(R)\rangle$, and that inferred from the KD reconstruction, $\langle \Sigma_{\rm gas,KD}(R)\rangle$.
\textbf{(b)} The relative bias, $\delta_\Sigma(R)$. 
\textbf{Alt Text}: Two line plots show azimuthally averaged gas surface-density profiles versus Galactocentric radius. Panel (a) compares the true simulation profile with the profile inferred from the KD reconstruction. Panel (b) shows the relative bias as a function of radius, highlighting systematic differences in the inner Milky Way.
}
\label{fig:SigmaProfiles}
\end{figure}

Figure~\ref{fig:SigmaProfiles}(a) compares the resulting profiles, $\langle\Sigma_{\rm gas,true}(R)\rangle$ and $\langle\Sigma_{\rm gas,KD}(R)\rangle$, for our fiducial model.
At large radii ($R\gtrsim 5$~kpc), the two profiles agree well, consistent with the relatively weak non-circular motions outside the bar region in our present setup.
In contrast, in the bar region ($R\simeq 0.2$--2~kpc) the true profile shows a clear depression, whereas the KD-reconstructed profile substantially fills in this dip and yields a much flatter radial trend.
This difference indicates that the axisymmetric KD inversion causes a net radial mixing in the inner Milky Way, smearing out the bar-induced deficit in the azimuthally averaged profile even though the KD face-on map simultaneously contains localized cavities and ridges.

To quantify the bias, Figure~\ref{fig:SigmaProfiles}(b) shows the relative radial difference
\begin{equation}
\delta_\Sigma(R) \equiv
\frac{\langle\Sigma_{\rm gas,KD}(R)\rangle - \langle\Sigma_{\rm gas,true}(R)\rangle}
     {\langle\Sigma_{\rm gas,true}(R)\rangle}.
\end{equation}
In the bar region, $\delta_\Sigma(R)$ becomes positive over the radial range where the true profile exhibits the dip, demonstrating that the KD reconstruction systematically overestimates the surface density there by reallocating gas from other radii.
Outside the bar region ($R\gtrsim 5$~kpc), $\delta_\Sigma(R) \approx 0$.

Overall, these results show that even if the classical near--far ambiguity is removed (as we can do in the simulation), an axisymmetric KD inversion applied to barred gas flows can erase or substantially weaken a true inner depression in the radial surface-density profile.
This provides a direct interpretation for part of the dispersion among published inner-Milky Way gas profiles and motivates caution when using KD-based face-on maps to quantify the depth and extent of the bar gap.

\subsection{Effects of gas self-gravity}
\label{subsec:selfgrav}

\begin{figure*}
\begin{center}
\includegraphics[width=1.0\textwidth]{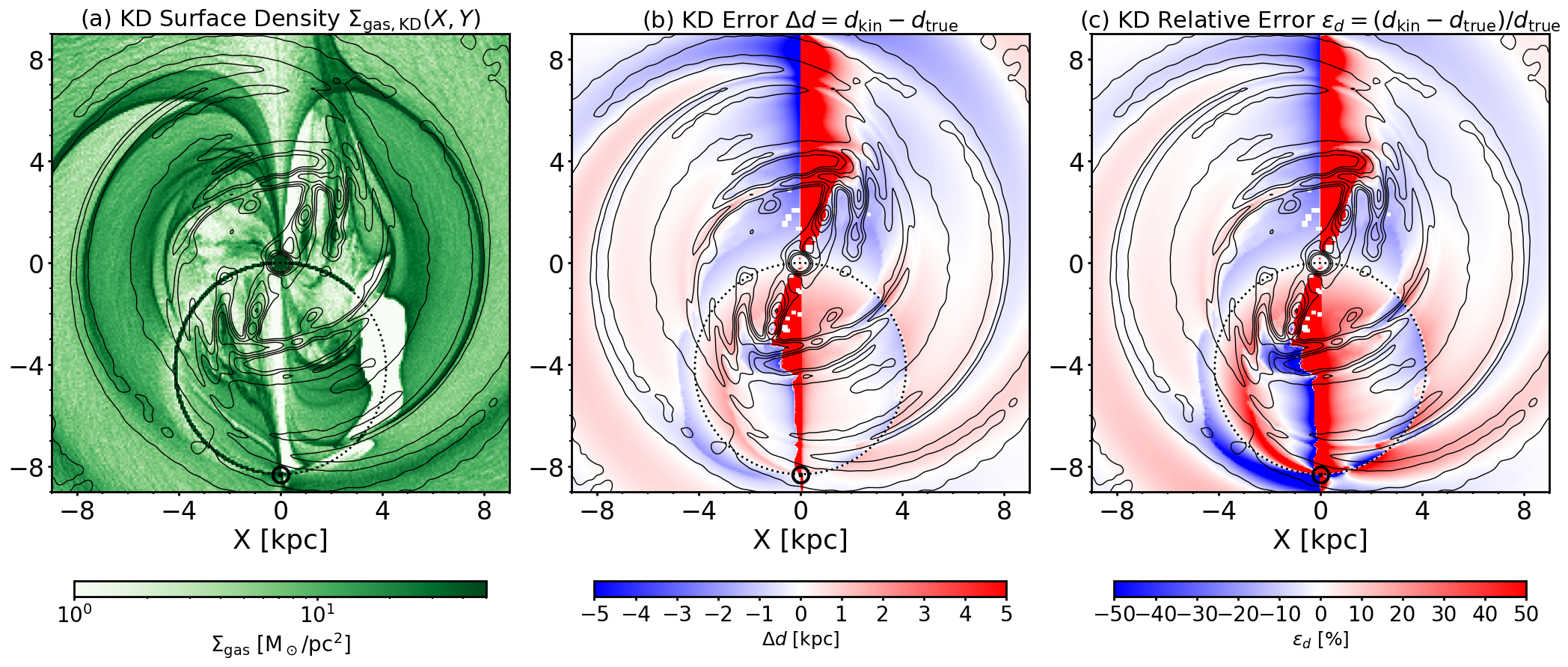}
\end{center}
\caption{
Same as Figure~\ref{fig:kd_map_v2}, but for the non-self-gravitating hydrodynamic simulation presented in \citet{Baba2025b}.
\textbf{Alt Text}: Three face-on maps in the Galactic plane, shown as in Figure 5, but for the non-self-gravitating hydrodynamic simulation. Panel (a) shows the KD reconstructed gas surface density with true-density contours. Panels (b) and (c) show the mean absolute and relative KD errors, allowing comparison of large-scale bias patterns with the fiducial self-gravitating model.
}
\label{fig:kd_map_nsg}
\end{figure*}

To test the robustness of our conclusions against small-scale turbulence and cloud formation, we repeat the KD reconstruction for a non-self-gravitating run and compare it to our fiducial self-gravitating model.
Figure~\ref{fig:kd_map_nsg} shows the KD reconstruction applied to the non-self-gravitating run presented in \citet{Baba2025b}.
In the absence of gas self-gravity, dense cloud-scale structures are strongly suppressed, the star-formation rate is lower, and stellar feedback is correspondingly weak. As a result, the velocity field is comparatively smooth, and the large-scale bar-driven non-circular streaming motions stand out more clearly than in our fiducial self-gravitating model (Figure~\ref{fig:kd_map_v2}).

By contrast, when gas self-gravity is included, local turbulent motions associated with self-gravitating clouds and feedback become non-negligible. In the KD reconstruction this enhanced small-scale velocity dispersion produces numerous line-of-sight elongated ``fingers-of-God'' features \citep[e.g.][]{Baba+2009}.
Despite these differences in the small-scale morphology, the KD-based face-on surface-density maps remain broadly similar between the two runs (Figure~\ref{fig:kd_map_nsg}a versus Figure~\ref{fig:kd_map_v2}a).
More importantly, the distance-error maps (panels~b and c) show that the large-scale pattern of systematic KD biases is nearly unchanged: outside of local fluctuations, the coherent regions of over- and under-estimated distances closely match between the self-gravitating and non-self-gravitating models.
This demonstrates that the dominant source of KD bias in the inner Milky Way is the global bar-driven streaming field rather than cloud-scale turbulence, while self-gravity primarily adds small-scale scatter that manifests as fingers-of-God artifacts in the reconstructed maps.

\subsection{KD reconstruction with a terminal-velocity rotation curve}
\label{sec:kd_obs_tvc}

In many observational KD applications, the circular-speed curve is inferred from the terminal velocity curve (TVC) under the assumption of purely circular rotation \citep[e.g.][]{Clemens1985,Sofue2017}, rather than computed from a mass model (equation~\ref{eq:vc}). To test whether our conclusions depend on this commonly adopted choice, we repeat the KD reconstruction using a TVC-based pseudo circular-speed curve, $V_{\rm c,ps}(R)$, obtained by converting the terminal velocities measured in the synthetic $(\ell,v)$ diagram via the standard terminal-velocity method. As shown by \citet[][their Fig.~3]{Baba2025b}, $V_{\rm c,ps}(R)$ exceeds the true (potential-based) rotation curve at $|\ell|\simeq 1^\circ$--$10^\circ$ ($R\simeq 0.2$--$1.5$~kpc) and peaks at $\sim 2\times$ the true rotation curve near $R\simeq 0.4$~kpc. This TVC-based curve is also consistent with the observational TVC-derived rotation curve of \citet{Sofue2017}.

\begin{figure*}
\begin{center}
\includegraphics[width=1.0\textwidth]{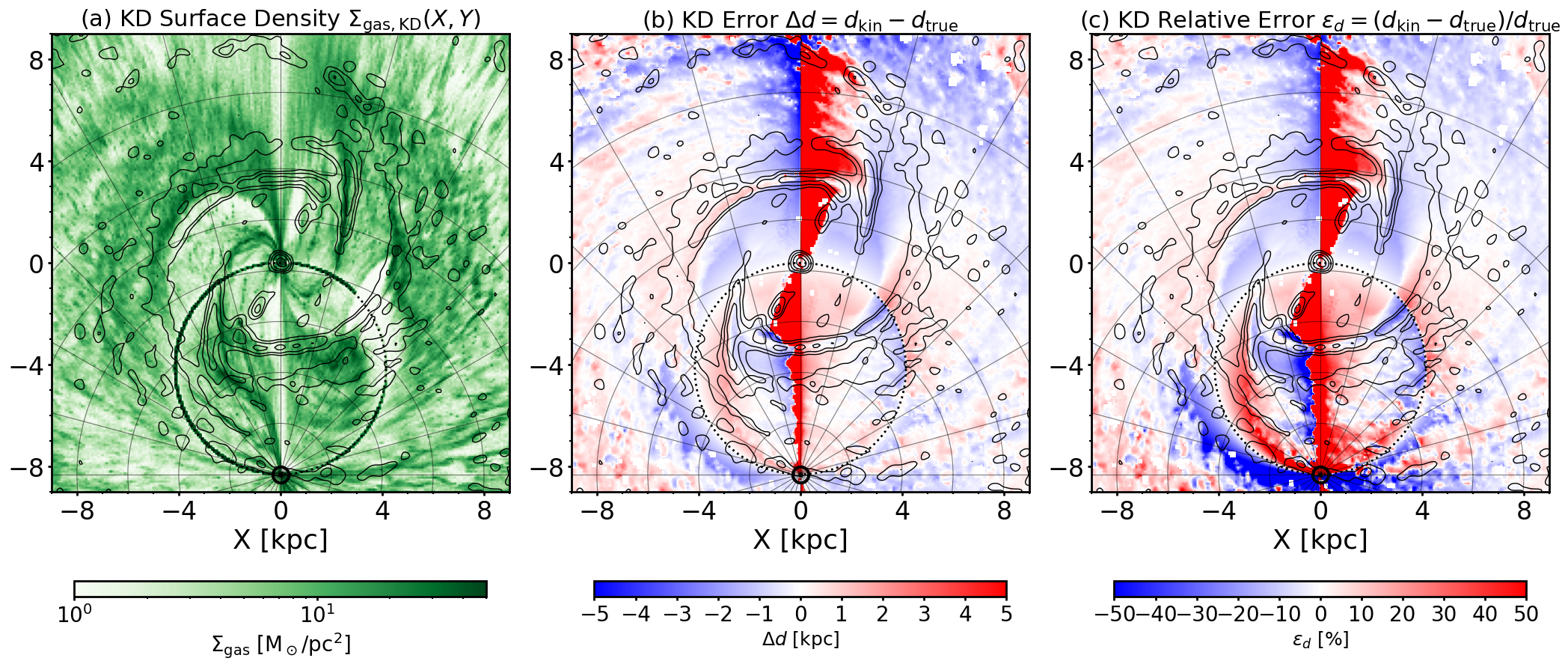}
\end{center}
\caption{
Same as Figure~\ref{fig:kd_map_v2}, but using the terminal-velocity-based ``pseudo'' circular-speed curve $V_{\rm c,ps}(R)$.
\textbf{Alt Text:} Same as Figure~\ref{fig:kd_map_v2}, but using the terminal-velocity-based pseudo circular-speed curve $V_{\rm c,ps}(R)$; the panels show the KD surface density, mean distance error, and mean relative distance error.
}
\label{fig:kd_map_tvc}
\end{figure*}

Figure~\ref{fig:kd_map_tvc} shows the KD reconstruction obtained with $V_{\rm c,ps}(R)$. The overall morphology of the KD artifacts is essentially unchanged relative to the fiducial reconstruction that uses the true (potential-based) rotation curve (Figure~\ref{fig:kd_map_v2}). 
This weak dependence on the adopted $V_{\rm c}(R)$ is expected from the first-order relation $\Delta d \simeq S\,\Delta V_{\rm LOS}$.
Although $V_{\rm c,ps}$ differs strongly from the true curve at $R\lesssim 1$~kpc, the KD sensitivity is small over much of the central region ($S\simeq 0$), except in the immediate vicinity of the tangent-point locus where $S$ formally diverges (Figures~\ref{fig:kd_method}b and \ref{fig:vlos_sens_model}b).
As a result, changes in $V_{\rm LOS}^{\rm circ}$ induced by adopting $V_{\rm c,ps}(R)$ translate only weakly into changes in the inferred distances away from the tangent points.
Consequently, the dominant spatial pattern of the KD errors remains set by the KD geometry (through $S$) and by bar-driven streaming, rather than by the specific choice of $V_{\rm c}(R)$.
We therefore conclude that the systematic KD failures in the bar region are driven primarily by bar-induced non-circular motions and the intrinsic KD geometry. Within the range motivated by terminal-velocity methods, the choice of circular-speed curve plays a secondary role.

\section{Summary}
\label{sec:Summary}

We have quantified how bar-driven non-circular motions bias Milky Way gas maps inferred with the classical KD method, using hydrodynamical simulations in an observationally constrained barred Milky Way potential. Applying a standard axisymmetric KD inversion to synthetic longitude--velocity data (while removing the near--far ambiguity to isolate non-circular effects), we find that the reconstruction is broadly reliable outside the bar-dominated region ($R \gtrsim 5$~kpc) in our present setup, but fails systematically in the bar region ($R \sim 0.5$--3~kpc). In the inner Milky Way, the KD face-on map develops highly anisotropic, quadrant-dependent artifacts, including arc-like over/underdense ridges and LOS-elongated low-density cavities. The associated distance-error maps show coherent structures with $|\Delta d|\sim 1$--2~kpc and relative errors of several tens of percent along the bar and inner ring, with sign reversals across the tangent-point locus and particularly large errors toward $\ell\simeq 0^\circ$. These patterns are captured to first order by $\Delta d \simeq S\,\Delta V_{\rm LOS}$, where the geometric KD sensitivity $S\equiv \partial d/\partial V_{\rm LOS}^{\rm circ}$ encodes the intrinsic ill-conditioning of the KD mapping \citep{Sofue2011}. When compressed into azimuthally averaged profiles, the anisotropic distance misassignments translate into a net radial redistribution of gas, substantially filling in the true bar-induced depression and yielding a much flatter inner profile. Together, these results imply that KD-based face-on maps and radial surface-density profiles in the barred inner Milky Way should be interpreted with great caution, even before considering additional observational systematics.

The impact of non-circular motions on KD has long been recognized, from spiral-arm streaming to bar-driven distortions of terminal velocities and inner rotation curves \citep[e.g.][]{Gomez2006,Hunter+2024,Baba2025b}. 
Our contribution is to provide a quantitative, spatially resolved calibration of where and by how much KD fails in a Milky Way bar potential constrained by stellar dynamical observations \citep[e.g.][]{Portail+2017,Sormani+2022agama}. 
Crucially, we connect the morphology of the KD errors to the intrinsic KD geometry via the sensitivity $S$, showing that the main sign patterns and amplitudes follow $\Delta d \simeq S\,\Delta V_{\rm LOS}$. 
This geometry-based view explains why KD reconstructions in the inner Milky Way can show strong quadrant dependence and why the inferred radial profiles can be systematically flattened by anisotropic distance misassignments.

Several simplifying assumptions should be kept in mind when applying our conclusions to the real Milky Way. We adopt a barred Milky Way potential and a fixed bar orientation relative to the Sun, chosen to be consistent with recent stellar-dynamical models and observational constraints. Although the main properties of the Galactic bar are now relatively well constrained \citep[e.g.][]{Portail+2017}, some quantitative uncertainty remains; alternative bar realizations would slightly modify the detailed streaming field and hence the fine structure of the KD bias pattern. Nevertheless, we expect our qualitative conclusion---that bar-driven streaming introduces strong, quadrant-dependent KD biases in the inner Milky Way---to be robust.

Motivated by the strong correlation between KD failures and the sensitivity $S$, a natural direction for future work is to develop a ``bar-informed KD'' method that replaces the purely circular-rotation assumption with a non-axisymmetric LOS-velocity model for the inner Milky Way. 
In practice, one can adopt
$V_{\rm LOS}^{\rm model}(\ell,b,d)=V_{\rm LOS}^{\rm circ}+\Delta V_{\rm LOS}^{\rm bar}(\ell,b,d)$,
where $\Delta V_{\rm LOS}^{\rm bar}$ is taken from gas-flow simulations in an observationally constrained Milky Way bar potential, and perform the KD inversion relative to this bar-informed velocity field to reduce systematic artifacts in the bar region. 
\citet{Pohl+2008} already explored a closely related, simulation-based approach in the context of CO deconvolution using barred gas-flow models, although with older bar potentials; applying the same idea with modern, stellar-dynamically constrained bar models would be a natural next step.

At the same time, both the standard and bar-informed KD methods share an intrinsic limitation: they infer the spatial distribution by mapping the observed LOS velocity field to distance. As a result, the recovered face-on map is not independent of the assumed velocity field, and the non-circular gas motions in the Milky Way cannot be determined independently of the inferred density distribution within the KD framework. To robustly constrain the true streaming field, it is therefore important to complement KD-based maps with approaches that retain and model the full kinematic information in $(\ell,b,v)$ space. A concrete example is the kinetic-tomography approach of \citet{Soler+2025}, which matches distance slices of a three-dimensional dust map to velocity channels of H\,\textsc{i} and CO cubes using a morphology-based statistic. This procedure yields a distance-resolved LOS velocity field and allows the streaming component to be quantified, rather than being absorbed into the distance inversion as in classical KD maps.

\section*{Funding}
This research was supported by the Japan Society for the Promotion of Science (JSPS) under Grant Numbers 21K03633, 21H00054, 22H01259, 24K07095, and 25H00394.

\begin{ack}
We sincerely thank the anonymous referee for their thoughtful and constructive comments, which helped improve the clarity and context of this paper.
Calculations, numerical analyses, and visualization were carried out on Cray XD2000 (ATERUI-III) and computers at the Center for Computational Astrophysics, National Astronomical Observatory of Japan (CfCA/NAOJ). 
\end{ack}

\appendix
\section*{Kinematic distance maps of CO gas}
\label{sec:kd_obs}

We construct a face-on map of the Galactic CO emission by assigning each $(\ell,b,v)$ voxel in the observed CO data cube \citep{Dame+2001}\footnote{Data available at \url{https://lweb.cfa.harvard.edu/rtdc/CO/CompositeSurveys/}.} to a Galactocentric position using a kinematic-distance model.
Our procedure largely follows the methodology developed by \citet{NakanishiSofue2003,NakanishiSofue2006,NakanishiSofue2016}, with minor modifications described below.
We assume axisymmetric circular rotation in the Galactic midplane and adopt the Solar parameters $(R_0,V_0)$.
For the rotation curve we use the compilation by \citet{Sofue2017}\footnote{Data available at \url{https://www.ioa.s.u-tokyo.ac.jp/~sofue/htdocs/2017paReview/}}.
Note, however, that in the bulge/bar region the \citet{Sofue2017} curve is largely constructed from terminal-velocity measurements of CO and H\,{\sc i}.
As pointed out by \citet{Baba2025b}, in the presence of strong non-circular streaming motions this terminal-velocity curve does not necessarily represent the true circular speed \citep[][]{Chemin+2015,Hunter+2024}.

For a trial heliocentric distance $s$ along a line of sight at longitude $\ell$ (we set $b=0$ for the kinematic inversion), the Galactocentric Cartesian coordinates are
\begin{equation}
x(s,\ell)=-s\sin \ell, \qquad y(s,\ell)=-R_0+s\cos \ell,
\end{equation}
and the Galactocentric radius is $R=\sqrt{x^2+y^2}$.
To avoid numerical issues at the origin, we impose a small floor $R\ge R_{\rm floor}$ (we use $R_{\rm floor}=10^{-3}\,{\rm kpc}$).
Given the adopted rotation curve $V_c(R)$, the circular velocity components in the inertial Galactocentric frame are
\begin{equation}
v_x = V_c(R)\,\frac{y}{R}, \qquad v_y = -V_c(R)\,\frac{x}{R}.
\end{equation}
The predicted line-of-sight velocity relative to the local standard of rest is then
\begin{equation}
v_{\rm los}(s;\ell) = (v_x+V_0)(-\sin \ell) + v_y \cos \ell,
\end{equation}
which defines a mapping $s\mapsto v_{\rm los}$ along each longitude.

For each longitude, we compute $v_{\rm los}(s;\ell)$ on a uniform distance grid $s\in[0.01,s_{\max}]$.
The tangent point is defined as the distance $s_{\rm tan}$ at which $v_{\rm los}$ reaches an extremum with the same sign as $\sin \ell$:
for $\sin \ell\ge 0$ (quadrants I--II) we take the maximum, while for $\sin \ell<0$ (quadrants III--IV) we take the minimum.
Observed velocities that exceed this envelope by more than $k_{\rm forb}\sigma_{\rm eff}$ are treated as ``forbidden'';
we adopt $\sigma_{\rm eff}=7~{\rm km\,s^{-1}}$ and $k_{\rm forb}=3$.
With the ``clip'' option used here, forbidden voxels are assigned to the tangent-point distance on the near branch and are ignored on the far branch.

Inside the Solar circle, an observed $v_{\rm los}$ generally corresponds to two heliocentric distances (near and far) on either side of the tangent point.
We compute these solutions by inverting $v_{\rm los}(s;\ell)$ separately on the two monotonic branches split at $s_{\rm tan}$.
For numerical stability, each branch is sorted by $v_{\rm los}$, duplicate velocities are removed, and the inverse relation $s(v_{\rm los})$ is obtained by linear interpolation.
This yields arrays $s_{\rm near}(v)$ and $s_{\rm far}(v)$ on the full observed velocity grid for each longitude.

To resolve the near--far ambiguity, we estimate the fractional contributions of the near and far solutions at each $(\ell,v)$ by fitting the observed latitude profile $T_b(b)$ over $|b|\le 5^\circ$.
For each $(\ell,v)$, we model the emission as a non-negative linear combination of two vertical components evaluated at the near and far distances,
\begin{equation}
T_b(b) \approx a_{\rm N}\,f\!\left[z_{\rm N}(b)\right] + a_{\rm F}\,f\!\left[z_{\rm F}(b)\right],
\end{equation}
where $z_{\rm N}(b)=s_{\rm near}\sin b$ and $z_{\rm F}(b)=s_{\rm far}\sin b$.
We adopt a $\mathrm{sech}^2$ vertical profile parameterized by a half-width $z_{1/2}$,
\begin{equation}
f(z) \propto \mathrm{sech}^2\!\left[ K\,\frac{z-z_0}{z_{1/2}} \right],
\qquad K=\ln(1+\sqrt{2}),
\end{equation}
and solve for the non-negative amplitudes $(a_{\rm N},a_{\rm F})$ using a non-negative least-squares fit.
The near and far weights are then defined as
\begin{equation}
w_{\rm N}=\frac{a_{\rm N}}{a_{\rm N}+a_{\rm F}}, \qquad
w_{\rm F}=\frac{a_{\rm F}}{a_{\rm N}+a_{\rm F}}.
\end{equation}
We adopt $z_{1/2}=0.06~{\rm kpc}$ for both the near and far components.
When the near and far kinematic solutions are too close to be separated reliably by the latitude profile, we set $(w_{\rm N},w_{\rm F})=(0.5,0.5)$.

For each $(\ell,b,v)$ voxel, we compute the velocity-integrated intensity $W_{\rm CO}=T_b\,\Delta v$ and convert it to an H$_2$ column density using a constant CO--H$_2$ conversion factor, $N({\rm H_2}) = X_{\rm CO}\,W_{\rm CO}$.
We adopt $X_{\rm CO}=2.0\times 10^{20}\ {\rm cm^{-2}\,(K\,km\,s^{-1})^{-1}}$ and include a factor 1.36 to account for helium.
The corresponding gas mass in the voxel is evaluated using the solid angle of the $(l,b)$ pixel and the distance-dependent area element.
We then deposit the voxel mass to the near and far $(x,y)$ locations implied by $s_{\rm near}(v)$ and $s_{\rm far}(v)$, weighted by $(w_{\rm N},w_{\rm F})$.
Because the sky footprint of a voxel maps to a distance-dependent area in the face-on plane, we accumulate both the deposited mass and the corresponding covered area, and define the face-on surface density as $\Sigma_{\rm H_2}=M/A$ on the final grid.


\end{document}